\documentclass[reprint,pra,showpacs,preprintnumbers,amsmath,amssymb,aps,superscriptaddress]{revtex4-1}
\usepackage{graphicx}
\usepackage{tikz}
\usepackage{subcaption}
\usepackage{dcolumn}
\usepackage{amsmath}
\usepackage{times}
\usepackage[braket, qm]{qcircuit}
\usepackage{physics}
\usepackage{mathtools}
\usepackage{wasysym}
\usepackage{caption}
\usepackage{booktabs}
\usepackage{multirow}
\usepackage{hhline}
\begin{document}

\title{Leakage mitigation for quantum error correction using a mixed qubit scheme}

\author{Natalie C. Brown}
\affiliation{School of Physics,
Georgia Institute of Technology, Atlanta, GA, USA}
\author{Kenneth R. Brown}
\affiliation{School of Physics,
Georgia Institute of Technology, Atlanta, GA, USA}
\affiliation{Schools of Chemistry and Biochemistry and Computational Science and Engineering, Georgia Institute of Technology, Atlanta, GA, USA}
\affiliation{Departments of Electrical and Computer Engineering, Chemistry and Physics, Duke University, Durham, NC, USA }

\date{\today}

\begin{abstract}

Leakage errors take qubits out of the computational subspace and will accumulate if not addressed.  A leaked qubit will reduce the effectiveness of quantum error correction protocols due to the cost of implementing leakage reduction circuits and the harm caused by interacting leaked states with qubit states. Ion trap qubits driven by Raman gates have a natural choice between qubits encoded in magnetically insensitive hyperfine states that can leak and qubits encoded in magnetically sensitive Zeeman states of the electron spin that cannot leak. In our previous work, we compared these two qubits in the context of the toric code with a depolarizing leakage error model and found that for magnetic field noise with a standard deviation less than 32 $\mu$G that the $^{174}$Yb$^+$ Zeeman qubit outperforms the $^{171}$Yb$^+$ hyperfine qubit.  Here we examine a physically motivated leakage error model based on ions interacting via the  M$\o$lmer-S$\o$renson gate.  We find that this greatly improves the performance of hyperfine qubits but the Zeeman qubits are more effective for magnetic field noise with a standard deviation less than 10 $\mu$G.  At these low magnetic fields, we find that the best choice is a mixed qubit scheme where the hyperfine qubits are the ancilla and the leakage is handled without the need of an additional leakage reduction circuit.



\end{abstract}
\maketitle
\section{Introduction}
\label{intro}
We have yet to discover the perfect qubit. Every known qubit candidate comes with assets and liabilities. Recently, there has been a growing interest in combining different qubit types in an effort to amplify these desirable attributes and suppress the undesirable noise. Such mixed qubit architectures look promising, addressing a wide range of issues such as cooling, crosstalk, and leakage \cite{inlek2017multispecies, tan2015multi, schaetz2015quantum,ballance2015hybrid, wang2017single, barrett2003sympathetic, negnevitsky2018repeated}. 

Qubits based on clock states are often favored in ion trapped quantum computers \cite{brown2016co, brown2011single, ballance2016high, blinov2004quantum}. Hyperfine qubits based on clock transitions suffer from virtually no memory errors because clock states
have a second-order dependence on magnetic field. However there exist additional energy states, resulting from Zeeman splittings, that are outside the defined computational subspace. These energy states can be accessed through leakage errors.  

Leakage errors are especially damaging. If left untreated they corrupt data and render error correction syndromes useless. Even so, standard error correction schemes are not adept to handle such errors. Additional leakage reducing circuits (LRC) are required to convert leakage errors into Pauli errors before they can be corrected \cite{aliferis2005fault, fowler, suchara}. 

Zeeman qubits are also a viable candidate for ion trap quantum computing \cite{lucas2004isotope, keselman2011high, poschinger2009coherent, ratschbacher2013decoherence}. While they suffer from a first-order dependence on magnetic fields and thus have more dephasing noise than hyperfine qubits, they have no additional energy states that lead to leakage. Thus the tradeoff is clear: they suffer more memory errors but do not suffer from leakage errors. 

In our previous work \cite{brown2018comparing}, we studied two specific types of qubits: $^{171}$Yb$^+$ hyperfine qubits and $^{174}$Yb$^+$ Zeeman qubits. We assessed the performance of a surface code built on each type of qubit, comparing the two different error models: one with leakage but no memory errors (hyperfine) and one with large memory errors but no leakage (Zeeman). We found that in certain magnetic field regimes, the Zeeman qubit's memory error can be suppressed enough that a surface code built on this type of qubit, outperforms one built on a hyperfine system.  

In this work, we study the performance of the surface code on a mixed qubit platform. Using $^{171}$Yb$^+$ hyperfine qubits for our ancilla and $^{174}$Yb$^+$ Zeeman qubits for our data, we reduce the potential for leakage errors at the cost of increasing memory errors. We simulate two different leakage models: a worst case stochastic model in which leaked qubits completely depolarize unleaked qubits they interact with and a M$\o$lmer-S$\o$renson model which captures the effects of leakage during a M$\o$lmer-S$\o$renson gate.  We find that in certain magnetic field regimes there is an improvement in the logical error rate of the surface code compared to the performance on either a pure hyperfine or Zeeman system. A surface code built on the mixed qubit architecture can effectively handle leakage without the use of a LRC. 

\section{Atomic Structure of Yb isotopes}
At the root of this study, we are investing the performance of the surface code when the ancilla and the data qubits have two very different error models.
 Since much of our work relies on errors that are very specific to the atomic structure of the qubits involved, we briefly outline the atomic structure of the two Ytterbium isotopes used in our simulation: $^{171}$Yb$^{+}$ ($I=1/2$) and $^{174}$Yb$^{+}$ ($I=0$).

The half integer nuclear spin of $^{171}$Yb$^{+}$ gives the well known hyperfine splitting responsible for the clock states $\ket{ F = 0 , m_F = 0 }$ and  $\ket{ F = 1 , m_F = 0 }$ that are often used to define a qubit. Because of their second-order magnetic field dependence, qubits based on this transition have virtually no idle errors. At a finite magnetic field with a stability of $10$ 
$\mu$G, the probability of a phase error for a hyperfine qubit is $10^{-15}$ smaller than the probability of a phase error for a Zeeman qubit \cite{brown2018comparing}. However, there exists additional energy states resulting from the Zeeman effect, $\ket{ F = 1 , m_F = -1 }$ and $\ket{ F = 1 , m_F = +1 }$. So the computational space defining the qubit is smaller than that of the physical system, leading to the possibility of leakage errors. $^{171}$Yb$^{+}$ is a good example to study since the leakage space is equal to the qubit space. The rate of leakage in and out of the computational space will then be equal. Other ions with a spin $1/2$ nucleus will also benefit from this symmetry (e.g. $^{133}$Ba$^+$ \cite{hucul2017spectroscopy}). Ions with larger spins, like $^{43}$Ca$^+$, will suffer from larger leakage rates due to the existence of a larger leakage space \cite{ballance2016high}. 

By contrast, $^{174}$Yb$^{+}$ has a zero nuclear spin. Thus the only energy states in the $S_{1/2}$ manifold are the two states resulting from Zeeman splitting ($\ket{ F = 1/2 , m_F = -1/2 }$ and  $\ket{ F = 1/2 , m_F = +1/2 }$) and it is these states that define the qubit. This is a double edge sword. On the one hand, since there are only two states available, there is no possibility for leakage. On the other, these states have a first-order dependence on magnetic field and thus will be highly susceptible to dephasing errors caused by fluctuations in the trap. 

It is worth noting that in each isotope, there exists higher-level leakage states in the D and F manifolds, but these states are quickly repumped back down to the ground state and are ignored in our analysis.

\section{Error Model}

\subsection{Sources of Physical Errors}
Raman transitions are a leading candidate for gate implementations in ion trap quantum computers. In the limit of no technical noise, the main source of error will arise from spontaneous scattering \cite{wineland, toolbox, ozeri2005hyperfine, uys, Ozeri}. While spontaneous scattering does not favor any particular state, the atomic structure  will affect how the scattering manifests. Raman scattering from these gates leaves the qubit in a different energy state. Depending on the atomic structure of the qubits, this leads to either Pauli $\hat{X}$ or $\hat{Y}$ type errors, or leakage errors. For hyperfine qubits, half of this scattering will result in leakage whilst for a Zeeman qubit, all the scattering results in Pauli type errors. Rayleigh scattering from these gates leaves the qubit in the same energy state but adds a phase. If the scattering from the two qubit levels is approximately equal, the scattering amplitudes can either destructively interfere leading to negligible errors (as is the case for $^{171}$Yb$^{+}$ ), or constructively interfere, leading to significant dephasing errors (as is the case for $^{174}$Yb$^{+}$ )  \cite{ozeri2005hyperfine, uys, Ozeri}. In the latter case, the probability of error resulting from Rayleigh scattering is approximately equal to that of Raman scattering. 

Another source of noise arises from magnetic field fluctuations in the trap. For the Zeeman qubit, the probability of error arising from the first-order effect grows quadratically with increasing field fluctuations. For the hyperfine qubit, the errors arising from the second-order effect grows quartically. For mean field fluctuations of higher than $10^{-4}$ G, the probability of error resulting from first-order effects is above 1$\%$,  the threshold error value of the surface code \cite{dennis2002topological, raussendorf2007fault, wang2009threshold}. However, even in these highly unstable fields, the probability of errors from the second-order effect is well below the threshold value \cite{brown2018comparing}. The noise resulting from these fields is significant for the Zeeman qubits and inconsequential for the hyperfine qubits.   
    
In our simulation, we vary the probability of scattering while considering a static error arising from the magnetic field. Based on the calculations of \cite{brown2018comparing}, we modeled the effects of scattering with the error channels: 
\begin{equation}
\mathcal{E}_{h}(\rho) = (1-\frac{p_s}{2})\rho+\frac{p_s}{8}\hat{X}\rho \hat{X}+\frac{p_s}{8}\hat{Y}\rho \hat{Y} +\frac{p_s}{4}\hat{L}\rho \hat{L} 
\end{equation}
\begin{equation}
\mathcal{E}_{Z}(\rho) =  (1-p_s)\rho+\frac{p_s}{4}\hat{X}\rho \hat{X}+\frac{p_s}{4}\hat{Y}\rho \hat{Y} +\frac{p_s}{2}\hat{Z}\rho \hat{Z}
\end{equation}
where $\mathcal{E}_{h}(\rho)$ and $\mathcal{E}_{Z}(\rho)$ is the error channel for the hyperfine and Zeeman qubits respectively and $p_s$ is the scattering error rate, and quantifies the overall error due to scattering. For the chosen detunings, the Raman and Rayleigh scattering lead to equal error on the Zeeman qubit but the hyperfine qubit experiences negligible decoherence due to Rayleigh scattering. This leads to hyperfine qubits having half the scattering error due to the qubit subspace occupying half the physical subspace. We expect for one-qubit gates $p_s$ = $9.76 \times 10^{-6}$ and two-qubit gates $p_s$ = $25.2 \times 10^{-5}$. For a more detailed discussion of how these errors manifest for the particular qubits, please refer to \cite{brown2018comparing}.



\subsection{Leakage Models}
While our error model is motivated by the physical error rates of the two ions considered, a more general view of our model is a system with one sided leakage. We defined one sided leakage as a system where only one qubit involved in a CNOT gate is able to leak. Because one-sided leakage could model the behavior of different physical systems other than ion traps, we looked at two different leakage models: depolarizing and M$\o$lmer-S$\o$renson.

The depolarizing leakage model has been used in numerous leakage simulations \cite{fowler, suchara, brown2018comparing, ghosh2013understanding}. In this model, when a leaked qubit interacts with an unleaked qubit via a CNOT, the leaked qubit remains in the leaked state while the latter is depolarized. This model is a worst-case stochastic model which may be applied to a variety of systems including superconductors \cite{ghosh2013understanding} and quantum dots \cite{fong2011universal, mehl2015fault}.

The second leakage model is specific to ion traps. It aims to capture the effect of how a leaked qubit interacts in a M$\o$lmer-S$\o$renson (MS) gate \cite{sorensen1999quantum}. The MS gate utilizes the motion of the ion crystal to couple the ions together. Two laser beams off resonantly detuned but close to the blue and red sidebands, drive the system causing both ions involved in the gate to change their state collectively \cite{haffner2008quantum, blatt2008entangled}. In a leaked ion, the spacing between the qubit energy state and the leakage energy state is large compared to the spacing between the collective motional modes of the crystal. This causes the lasers to be much farther off resonant and both on the same side of the carrier transition. The leaked ion will then only get weakly displaced. Thus when an MS gate is performed with a leaked ion, no entanglement is generated \cite{mike}.

The full CNOT gate involves several more single qubit gates that still get applied whether or not the MS gate failed \cite{maslov2017basic}. If the control leaked, the target undergoes a $X(-\pi/2)$ rotation. If the target leaked, the control undergoes a $Z(-\pi/2)$ rotation. We simulate this by applying a Pauli-twirl approximation which gives the channels:  
\begin{equation}
\mathcal{E}_{bit}(\rho) = \frac{1}{2}\rho +\frac{1}{2}\hat{X}\rho \hat{X} \\
\end{equation}
\begin{equation}
\mathcal{E}_{phase}(\rho) = \frac{1}{2}\rho +\frac{1}{2}\hat{Z}\rho \hat{Z} \\
\end{equation}
as used in \cite{mike}. We applied $\mathcal{E}_{bit}(\rho)$ to the target if the control has leaked and $\mathcal{E}_{phase}(\rho)$ to the control if the target has leaked. In our one sided leakage model this translates to applying $\mathcal{E}_{bit}(\rho)$ to our data qubits if an ancilla is leaked during our $X$ stabilizer syndrome extraction or $\mathcal{E}_{phase}(\rho)$ to the data if an ancilla leaked during our $Z$ stabilizer extraction. 

We make several assumptions in both our leakage models. First we assume that leakage is only caused by spontaneous scattering from the gates and thus initialization of the qubit does not cause leakage. Typically, ions are initialized using optical pumping techniques which do not result in leakage. This assumption has also been made in other leakage studies \cite{fowler}. Second, we assume that a leaked qubit has a probability to return to the computational subspace equal to the probability that it leaked out. This is again motivated by physical scattering events and has been modeled in several other studies \cite{brown2018comparing, fowler, suchara, mike}. Finally, we assume a leaked qubit remains leaked until it leaks back to the computational space or is reinitialized.     

\section{Surface code Simulation}
Topological surface codes are a leading candidate for quantum error correction (QEC), due to their high thresholds and single ancilla extraction \cite{trout2018simulating, PhysRevA.90.032326, dennis2002topological, kitaev2002classical, kitaev1997quantum}. However, standard topological codes alone are incapable of handling leakage errors. If left unhandled, the performance of the surface code suffers dramatically from the correlated errors produced from a single leakage error \cite{fowler, suchara, brown2018comparing, aliferis2005fault, mike}. Fortunately, Alferis and Terhal showed that a threshold exists for the surface code in the presence of leakage if one incorporates leakage reducing circuits (LRC) \cite{aliferis2005fault}. Typically, these LRC's involve teleporting or swapping leaked qubits with an auxiliary qubit. While LRC's are effective for reducing leakage, they come at a cost. Implementing even the simplest LRC involves incorporating more gates and thus adds more potential fault locations. 

\begin{figure}[h]
\includegraphics[trim=250 460 250 125, width=4cm, height=7cm]{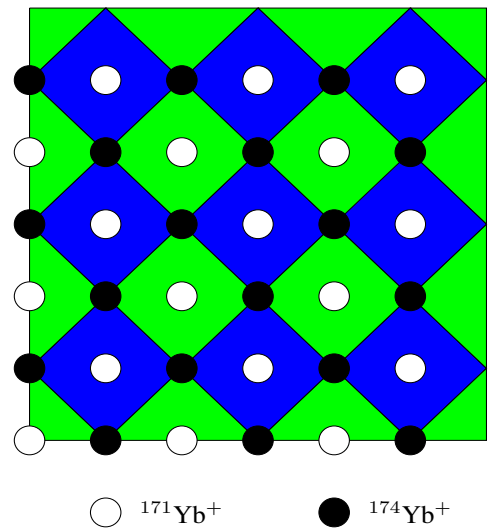}
\caption{Mixed species surface code layout. Hyperfine ($^{171}$Yb$^+$) ions are defined as ancilla qubits (white) and Zeeman ($^{174}$Yb$^+$) ions are defined as data qubits (black). The green (blue) diamond represents an X (Z) stabilizer measurement.}
\label{YbModelSurf}
\end{figure}

\begin{figure}[h]
\includegraphics[trim=250 475 250 150, width=4cm, height=5.5cm]{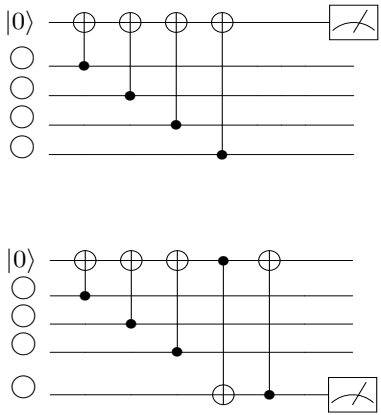}
\caption{The top circuit is the standard syndrome extraction circuit for the surface code. The bottom circuit is the standard circuit with a SWAP-LRC implemented at the end. Both the homogenous Zeeman system and the mixed species system utilize the top circuit. The homogenous hyperfine system requires the LRC to handle leakage errors.}
\label{circuits}
\end{figure}

In a surface code, qubits are arranged on a lattice and function either as data qubits, which encode the information, or ancilla qubits, which are used to measure error syndromes (see Fig. \ref{YbModelSurf}). In the standard surface code, syndrome extraction is accomplished by performing four CNOT gates between each data and ancilla qubit  and then measuring the ancilla (see Fig. \ref{circuits}). Minimum weight perfect matching is done to infer the most probable error given the observed syndrome. 

In a surface code built on only Zeeman qubits, this standard syndrome extraction is all that is needed to detect and correct errors. In a surface code built on only hyperfine qubits, a LRC must be implemented to convert leakage errors into Pauli errors. 

The simplest method for implementing a LRC is to add a SWAP gate at the end of the syndrome extraction circuit. Data and ancilla qubits swap their roles and thus leaked qubits get reinitialized at most every other cycle. Thanks to gate identities, this amounts to adding one extra CNOT to the error correction cycle (see Fig. \ref{circuits}). We refer to this LRC as the SWAP-LRC.  

In our simulation, we assign the role of data to the $^{174}$Yb$^+$ Zeeman qubits and the role of ancilla to $^{171}$Yb$^+$ hyperfine qubits. Since data qubits cannot leak, there is no need for a LRC. In fact, leaked qubits can live for at most one error correction cycle. This is already an improvement over the pure hyperfine system where leaked qubits can live twice as long. 

Furthermore, when a leaked qubit enters a CNOT gate, the other qubit involved incurs some error, as dictated by the errors models discussed above. For a pure hyperfine system there are potentially four such corrupt gates, because data qubits can leak and leakage is not necessarily eliminated every cycle. For the mixed species system, there are only three such corrupt gates since only ancilla can leak and we assumed initialization does not cause leakage. 

While the advantages of the mixed species system over a hyperfine system are immediately clear, they come at a cost. While we no longer require a LRC to handle leakage errors, we have effectively traded in half our leakage errors, which vary with the scattering rate, for constant memory errors. Still, memory errors manifest as Pauli $\hat{Z}$ errors which we can correct without additional overhead and, compared to a pure Zeeman system, the mixed species system will incur half the memory errors due to the symmetry of the surface code.

\section{Results and Discussion}

Implementing the two different error models for the hyperfine and Zeeman qubits discussed above, we examined the performance of the surface code built on this mixed species structure and compared it to the performance of the pure hyperfine and Zeeman systems. In each simulation, we varied the probability of a spontaneous scattering event ($p_s$) while applying a constant magnetic field error probability ($p_M$).  We simulated the effects of both the depolarizing and MS leakage models and looked at a range of magnetic field stabilities (see Table \ref{table1}) to get a grasp on where the trade off between leakage errors and memory errors might lie.

\begin{figure}[h]
\includegraphics[width=8cm, height=5cm]{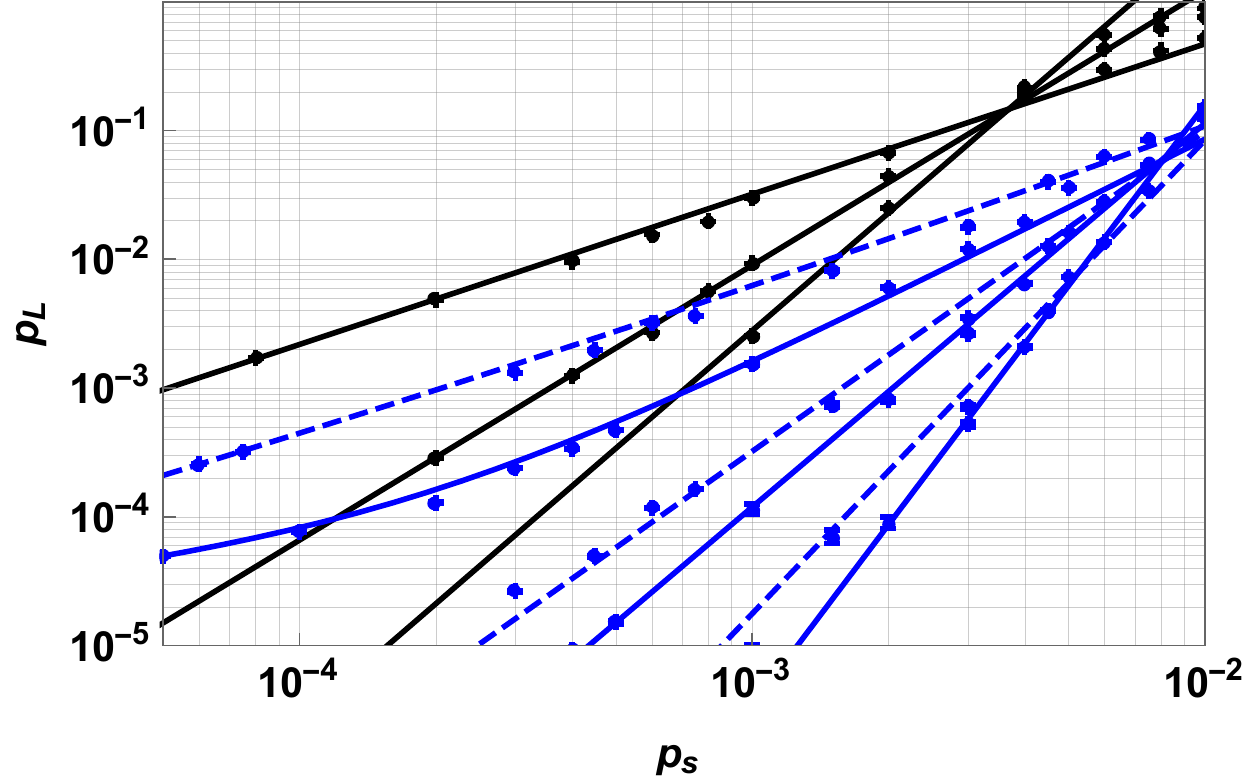}
\caption{Distance comparison for the depolarizing leakage model of distance 3, 5, 7. The solid and dashed blue lines represent the Zeeman and mixed species systems respectively, stabilized to $10$ $\mu$G standard deviation from the mean magnetic field per two qubit gate. The solid black line represents the hyperfine system with the SWAP-LRC implemented. The logical error rate ($p_L$) is proportional to $p_s^{\lceil \frac{d}{2}\rceil}$ for the Zeeman system and $p_s^{\lceil \frac{d}{4}\rceil}$ for mixed species, and hyperfine systems.  }
\label{dist_comp_DP}
\end{figure}

\begin{figure}[h]
\includegraphics[width=8cm, height=5cm]{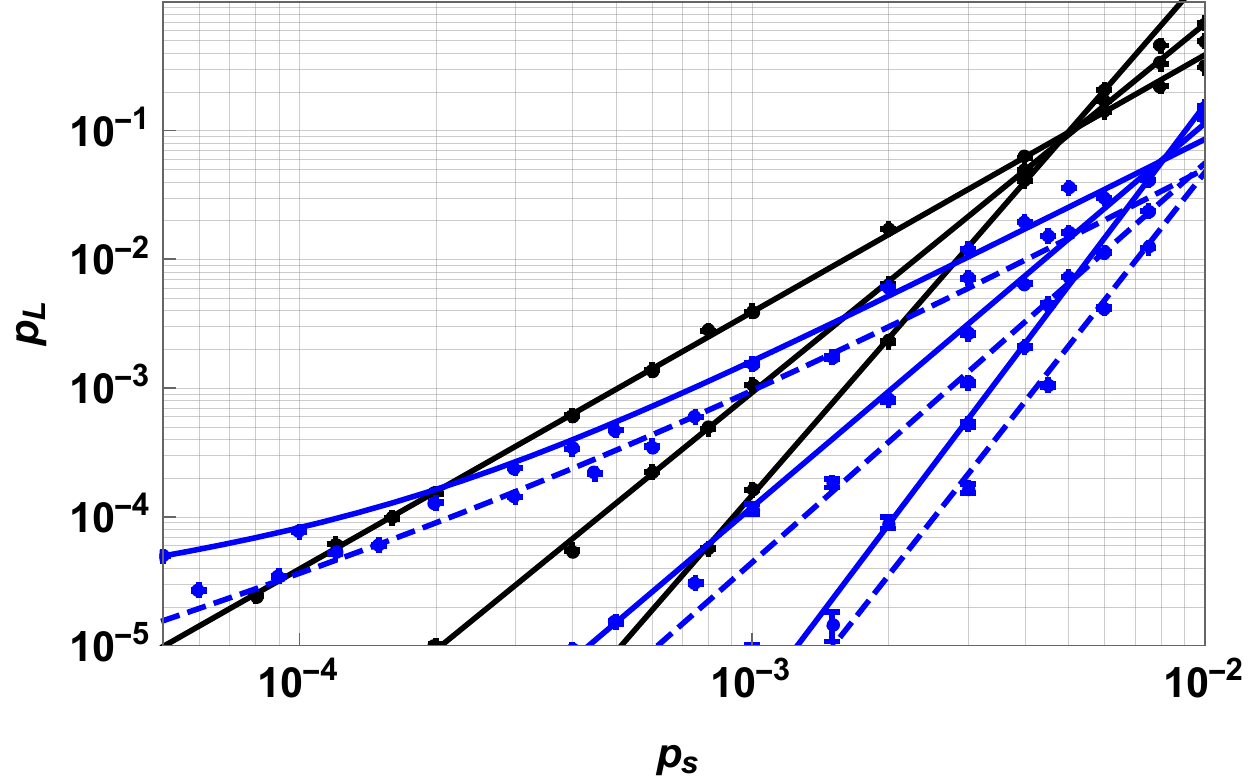}
\caption{Distance comparison for the MS leakage model of distance 3, 5, 7. The solid and dashed blue lines represent the Zeeman and mixed species systems respectively, stabilized to $10$ $\mu$G standard deviation from the mean magnetic field per two qubit gate. The solid black line represents the hyperfine system with the SWAP-LRC implemented. The logical error rate ($p_L$) is proportional to $p_s^{\lceil \frac{d}{2}\rceil}$ for the Zeeman, mixed species, and hyperfine systems.   }
\label{dist_comp_MS}
\end{figure}

\subsection{Leakage effects}
In the depolarizing leakage model, a single leakage error on an ancilla can cause a two-qubit error chain by depolarizing its neighboring data qubits. These hook errors reduce the code's effective distance by half  \cite{fowler, mike}, see Fig. \ref{dist_comp_DP}. In the hyperfine system, these ancilla qubits then get swapped and reassigned as data qubits. Leaked data qubits will corrupt ancilla qubits, leading to measurement errors.  

In the MS leakage model, leakage errors on ancilla can cause errors on data qubits that are of the same type as that stabilizer. All potential error outcomes are either a single-qubit or two-qubit error, up to a stabilizer. Thus any $\lfloor \frac{d-1}{2} \rfloor$ physical error does not produce a logical error and the effective code distance is maintained, see Fig. \ref{dist_comp_MS}. Leakage errors on data can produce many time correlated errors but they will not produce any additional space correlated errors since a leaked data qubit cannot spread errors to ancilla that will then propagate to other data qubits \cite{mike}.  

In the mixed species system, there are less time and space correlated errors for two reasons: leakage can only live for one cycle, and leaked ancilla never get swapped with data qubits. This is independent of the leakage error model used. So we expect the mixed species model to outperform the hyperfine system if the memory errors can be suppressed. 

For the depolarizing model (Fig. \ref{dist_comp_DP}), ancilla leakage is so damaging that the mixed species system, even with half the potential for leakage errors, suffers a logical error rate suppression proportional to $\lceil \frac{d}{4} \rceil$ log($p$). While in certain magnetic field regimes this removal of potential leakage errors is enough to beat the hyperfine system, the mixed species model will \textit{almost} never be able to do better than the Zeeman system in the same error regime. Having half the memory errors is not enough to compensate for the damage leakage can cause. Of course, this all rests on the effects from the magnetic field, which we will discuss in detail later.  

For the MS model (Fig. \ref{dist_comp_MS}), leakage is much less damaging and we see every system behaves fault tolerantly. In this leakage model, the mixed species system has the lowest logical error rate. It beats the hyperfine system for the same reasons as the depolarizing model (i.e. less leakage and shorter lived leakage) and it beats the Zeeman system because the structure of the leakage errors imposed by the MS model makes leakage errors more comparable to memory errors. In fact, leakage errors are less damaging than two-qubit dephasing errors. While they cause errors on other qubits, the structure of the MS leakage model restricts these errors to be the same as the stabilizer. In the Zeeman model, this is not true for all ancilla; $Z$ type ancilla will have this advantage but for $X$ type ancilla, dephasing errors will cause measurement errors. Because the mixed species system suffers less of these dephasing errors, in no magnetic field regime will the pure Zeeman system outperform the mixed species system.

\subsection{Memory effects}
For both leakage models, when the main source of error arises from spontaneous scattering ($p_s > p_M$), we see an improvement in the logical error rate as the scattering probability decreases. Once the scattering rate decreases below the static memory error probability ($p_s < p_M$), the logical rate rate plateaus as memory errors dominate. The hyperfine system is immune to these memory errors and so its performance is the same for every magnetic field stability. 
Table \ref{table1} lists the values of the static $p_M$ applied in our simulation.  

\begin{figure}[h]
\includegraphics[width=8cm, height=5cm]{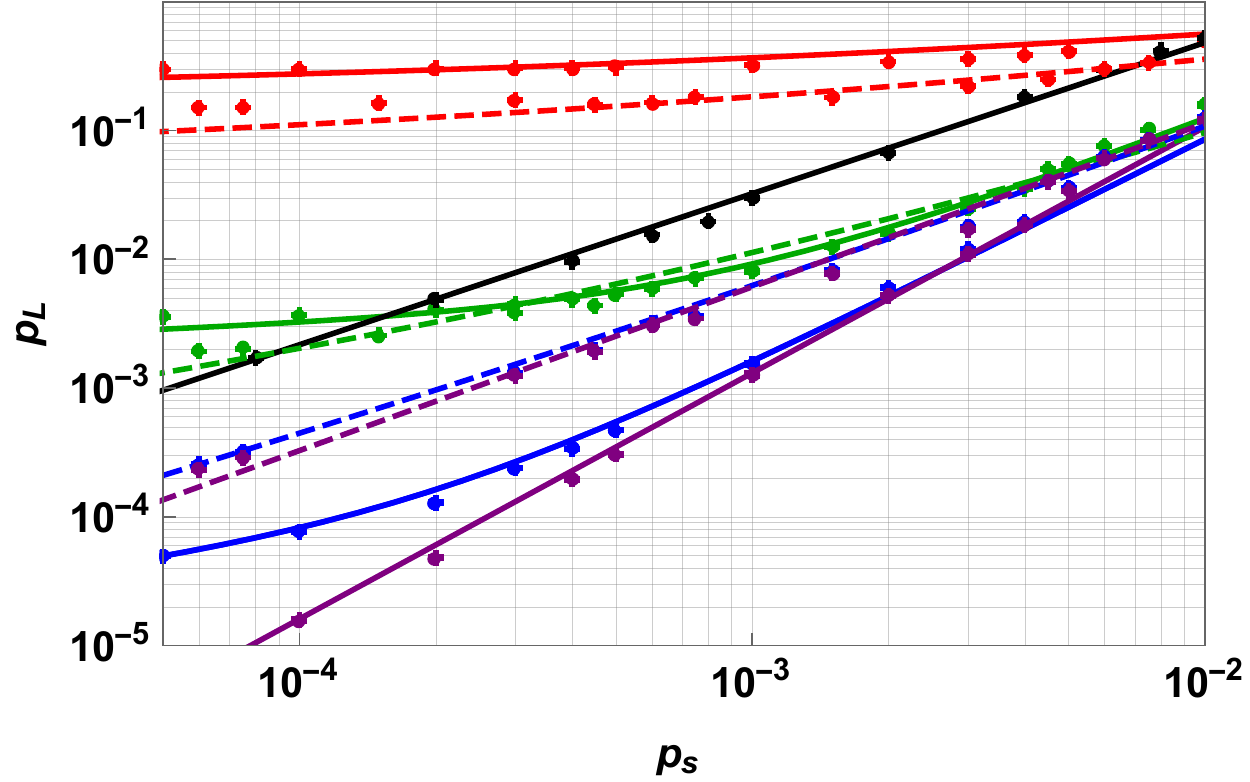}
\caption{Comparison of the different schemes for a distance-3 surface code using the depolarizing leakage model. The solid and dashed colored lines represent the Zeeman and mixed species  systems respectively. The solid black line shows the performance of the hyperfine system with the SWAP-LRC implemented.  The color of the line indicates the standard deviation from the mean magnetic field per two qubit gate: $100$ $\mu$G (red), $32$ $\mu$G (green), $10$ $\mu$G (blue) and $1$ $\mu$G (purple).}
\label{d=3_DP}
\end{figure}

\begin{figure}[h]
\includegraphics[width=8cm, height=5cm]{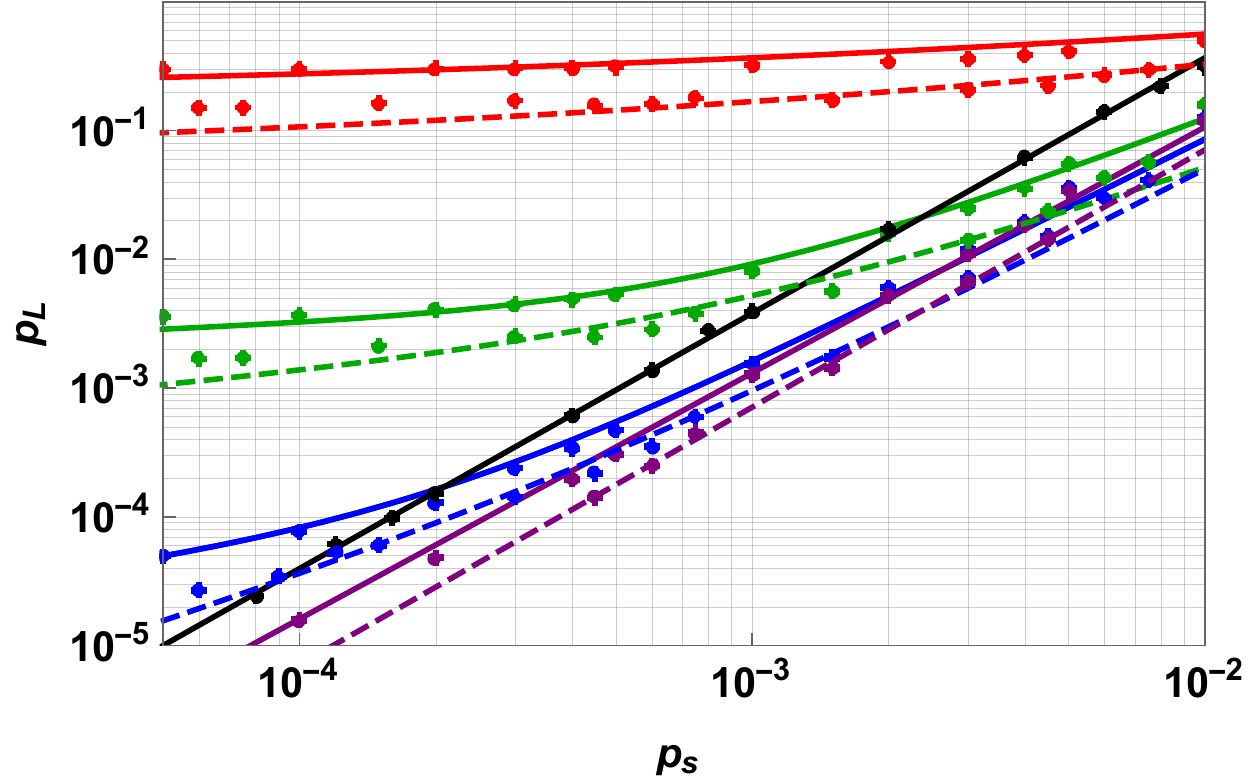}
\caption{Comparison of the different schemes for a distance-3 surface code using the MS leakage model. The solid and dashed colored lines represent the Zeeman and mixed species systems respectively. The solid black line shows the performance of the hyperfine system with the SWAP-LRC.  The color of the line indicates the standard deviation from the mean magnetic field per two qubit gate: $100$ $\mu$G (red), $32$ $\mu$G (green), $10$ $\mu$G (blue) and $1$ $\mu$G (purple).}
\label{d=3_MS}
\end{figure}

\begin{center}
\begin{table}
\begin{tabular}{  | c | c | }
  \hline
  S. D. ($\mu$G) & $p_M$ \\ \hhline{|=|=|}
  $\sigma=100$ & $7.75 \times 10^{-3}$ \\ \hline 
  $\sigma=32$ & $7.75 \times 10^{-4}$ \\ \hline
  $\sigma=10$ & $7.75 \times 10^{-5}$ \\ \hline
  $\sigma=1$ & $7.75 \times 10^{-6}$ \\ 
  \hline
\end{tabular}
\caption{A list of error probabilities caused by the first-order Zeeman effect ($^{174}$Yb$^+$). $\sigma$ is the standard deviation from the mean magnetic field per two qubit gate in $\mu$G.}
\label{table1}
\end{table}
\end{center}

The logical error rate of a distance-3 surface code using the depolarizing leakage model can be seen in Fig. \ref{d=3_DP}. When $p_M > p_s$, then the performance of the surface code is limited by the amount of memory errors incurred. Since the Zeeman system suffers the most from these errors, it has the worst logical error rate of the three systems. The mixed species suffers half as many memory errors and thus will always be better the Zeeman system but worst than the hyperfine system in \textit{most} of this regime.  

When $p_M < p_s$, the performance of the surface code is limited by the amount of leakage incurred. Since the hyperfine system suffers the most from leakage, it has the worst logical error rate. The mixed species code will always be better the the hyperfine system but always worst than the Zeeman system in this regime.  

There is a small range when $p_M > p_s$ in which the mixed systems system has the lowest logical error rate. In the depolarizing leakage model, leakage errors cause more damage than memory errors. The hyperfine system not only has more potential for leakage, it also has more fault locations due to the extra gate needed for the SWAP-LRC. There is a small range for $p_s$, when the total probability of a logical error caused from a single leakage event in the hyperfine system is \textit{higher} than the probability of a logical error caused by either a single leakage or two memory errors in the mixed species system. When this is true, the mixed species system outperforms the hyperfine system.

The logical error rate of a distance-3 surface code using the MS leakage model can be seen in Fig. \ref{d=3_MS}. In this leakage model, memory errors are more damaging than leakage errors. Thus there is no magnetic field regime in which the pure Zeeman system will outperform the mixed species system. When $p_M > p_s$, the hyperfine system will have the lowest logical error rate. 

In fact, we have the opposite situation of the depolarizing model: there is a small regime when $p_s > p_M$, in which the probability of a logical error caused from two leakage errors in the hyperfine system is \textit{lower} than the probability of a logical error caused by two leakage errors or two memory errors in the mixed species system. Since memory errors are more damaging, the stability of the magnetic field required to suppress the memory errors in order to see an advantage in using a Zeeman qubit is higher than when compared to the depolarizing leakage model. For the errors we are interested in, the magnetic field stability for the Zeeman qubits becomes stricter than our previous estimates with this error model \cite{brown2018comparing}.


For the ions considered, the total scattering probability for a two qubit gate was calculated to be $2.5 \times 10^{-4}$ \cite{brown2018comparing}. In these calculations we assumed the gates were driven by co-propagating linearly polarized Raman beams with a laser frequency of $355$ nm and a two qubit gate time of $200$ $\mu$s. These parameters minimize spontaneous scattering and reflect parameters used in recent experiments \cite{linke2017fault, leung2018robust, debnath2016demonstration, fallek2016transport}. 

For this realistic total scattering probability ($p_s = 2.5 \times 10^{-4}$), in each leakage model there is a magnetic field regime where the mixed species outperforms both homogenous systems. For the depolarizing model, we can see this is a narrow window near a stability of 32 $\mu$G. Below this value, the homogenous Zeeman qubit yields better performance. For the M$\o$lmer-S$\o$renson leakage model, leakage is less damaging and a lower memory error is required to outperform the homogenous hyperfine qubit. Below 10 $\mu$G the Zeeman and mixed species system outperform the pure hyperfine system with the mixed species providing a fractional improvement over the Zeeman system corresponding to $1/2$ and primarily due to hyperfine qubits having a lower overall error rate after scattering than Zeeman qubits.  Zeeman qubits have already been realized in fields stabilized to 10 nG, well below either model's requirement \cite{ruster2016long}.

\section{Conclusions}
In this work we have shown an advantage of mixing qubit types together in order to limit the effects of leakage. The advantage of using mixed-species depends on the details of how leaked qubits interact with qubits in the computational subspace. There are other advantages that a mixed species platform could provide. 

In our simulations we did not take into account different state preparation and measurement errors (SPAM) associated with the two different types of qubits. Hyperfine qubits typically have less SPAM errors as they can be easily measured reliably using state selective fluorescence 
\cite{crain2019high, noek2013high}. For the typical magnetic field strengths used in ion trap quantum computing, the frequency separation between the Zeeman qubits states (typically $8.2$ - $20$ MHz) is smaller than the natural $P$ level spectral width of $19.6$ MHz \cite{PhysRevA.82.063419}. State selective fluorescence cannot be directly applied in this case and the qubit must be first shelved to a different energy level before it can be measured \cite{toolbox}. In our mixed species scheme, the qubits that get measured often (ancilla) correspond to the qubits that are easy to measure (hyperfine). 

Another intrinsic advantage of the mixed species system is its ability to limit crosstalk. Because the qubits are no longer identical, laser spillage on adjacent ions can no longer be a problem. Here the isotopic separation only reduces crosstalk but by using distinct species (e.g Be$^+$ and Ca$^+$ \cite{ballance2015hybrid, negnevitsky2018repeated}, Mg$^+$ and Be$^+$ \cite{barrett2003sympathetic}, Yb$^+$ and Ba$^+$ \cite{wang2017single}) crosstalk could be eliminated. Mixed species systems could also help with cooling issues by allowing Doppler cooling without damaging the data.  

Our results also emphasize the importance of leakage models. For the depolarizing leakage model, leaked ancilla are so damaging, a single physical leakage error leads to a logical error. For the MS leakage model, the Pauli twirl approximation gives a convenient result that makes ancilla leakage less dangerous than stochastic Pauli errors. The way in which leakage is modeled can also determine which surface code is  best suited to handle the correlated errors associated with leakage \cite{mike}. 

Our results show that the Zeeman and mixed species systems will outperform the homogenous hyperfine system for stable magnetic fields.  For the depolarizing leakage error model and stable magnetic fields. The homogenous Zeeman system outperforms the mixed species systems except for a small region of parameter space. For the MS leakage error model, the magnetic field must be more stable, but then the mixed species systems outperforms the Zeeman system for all scattering error rates. 

These results highlight the fact that ancilla leakage is more dangerous than data leakage. It is natural to wonder why we used hyperfine qubits as ancilla and Zeeman as data and not the other way around. While ancilla leakage is more damaging, the standard error correction circuit naturally removes leakage without the need to implement any LRCs. If data leaks, while it might not be as damaging in any given error correction cycle, something \textit{must} be done to remove it or else it will continued to wreak havoc. At the circuit level, this means implementing an LRC. Adding a SWAP-LRC at the end to reduce the data leakage would mean the following error correction round would result in leaky ancilla. Reversing the roles of the hyperfine and Zeeman qubits not only requires additional gates for the LRC, it would also result in leakage living twice as long. Leaked ancilla would be able to live on as leaked data before being removed. 

The periodic boundary conditions of the toric code help with the implementation of the SWAP-LRC. The periodicity guarantees every qubit will have a qubit to swap with at the end of the cycle. While such boundary conditions could be implemented on modular architectures \cite{NickersonNatComm2013} and single ion chains \cite{trout2018simulating}, the mixed species system is not restricted by these boundary conditions and could be easily implemented on any planar architecture suited for the surface code  \cite{LekitscheSciAdv2017}. To implement the SWAP-LRC on a plane, additional qubits could be added to the boundary and swapped up and down every other cycle \cite{suchara, ghosh2015leakage} 

In our study, we did not consider any other LRC implementations. We choose to look at the SWAP-LRC since it requires the least amount of overhead. We also did not consider any physical methods for leakage removal, which could in practice remove populations from the leaked qubit state. Our aim was to demonstrate the effectiveness of a surface code with leakage errors but no LRCs. 

Leakage errors are a fundamental limiting error in ion trap quantum computers made with hyperfine qubits. Even in systems built on microwave gates \cite{harty2016high, lekitsch2017blueprint, ospelkaus2011microwave}, which do no suffer from the spontaneous scattering effects, background gas collisions can cause leakage. Leakage is a damaging error that needs special consideration when designing new systems. 

Memory errors are also a limiting error but pose more a technical challenge. Improvements in field stability will further suppress the rate of memory errors incurred on a system. This is an active area of research where magnetic field stability continues to improve \cite{PhysRevX.7.031050}.   

For near term experiments, we do not anticipate leakage being the main source of error. The probability of leakage errors is low. There is more technical noise to overcome before we see the effects of leakage dominate. But when constructing large scale fault tolerant devices, we must consider the tradeoffs between overhead of handling such errors and mitigating their effects through design. We expect to see many other advantages for mixing qubit types in the future.

\section{Acknowledgments}
We thank Michael Newman for useful discussions on the effects of leakage and Muyuan Li and Dripto Deboy for insights on gate errors and surface code simulation. We also thank Andrew Cross and Martin Suchara for providing the toric code simulator, with permission from IBM. This work was supported by the Office of the Director of National Intelligence - Intelligence Advanced Research Projects Activity through Army Research Office (ARO) contract W911NF-10-1-0231, ARO MURI on Modular Quantum Systems W911NF-16-1-0349, and EPiQC - a National Science Foundation Expedition in Computing 1730104.

\bibliography{inversion}

\end{document}